\documentclass{article}
\usepackage{amssymb}

\input{tcilatex}
\begin{document}

\author{ Giuseppe Castagnoli \and Pieve Ligure (Genova)
giuseppe.castagnoli@gmail.com}
\title{The 50\% advanced information rule of the quantum algorithms}
\maketitle

\begin{abstract}
The oracle chooses a function out of a known set of functions and gives to
the player a black box that, given an argument, evaluates the function. The
player should find out a certain character of the function (e. g. its
period) through function evaluation. This is the typical problem addressed
by the quantum algorithms. In former theoretical work, we showed that\ a
quantum algorithm requires the number of function evaluations of a classical
algorithm that knows in advance 50\% of the information that specifies\ the
solution of the problem. This requires representing physically, besides the
solution algorithm, the oracle's choice.

Here we check that this \textit{50\% rule} holds for the main quantum
algorithms. In the structured problems, a classical algorithm with the
advanced information, to identify the missing information should perform one
function evaluation. The speed up is exponential since a classical algorithm
without advanced information should perform an exponential number of
function evaluations. In unstructured database search, a classical algorithm
that knows in advance 50\% of the $n$\ bits of the database location, to
identify the $n/2$ missing bits should perform $O\left( 2^{n/2}\right) $
function evaluations. The speed up is quadratic since a classical algorithm
without advanced information should perform $O\left( 2^{n}\right) $ function
evaluations. The 50\% rule identifies the problems solvable with a quantum
sped up by comparing two classical algorithms, with and without the advanced
information. The advanced information classical algorithm also defines the
quantum algorithm that solves the problem. Each classical history,
corresponding to a possible way of getting the advanced information and a
possible result of computing the missing information, is represented in
quantum notation as a sequence of sharp states. The sum of the histories
yields the function evaluation stage of the quantum algorithm. Function
evaluation entangles the oracle's choice register (containing the function
chosen by the oracle) and the solution register (in which to read the
solution at the end of the algorithm). Information about the oracle's choice
propagates from the former to the latter register. Then the basis of the
solution register should be rotated to make this information readable. This
defines the quantum algorithm, or its iterate and the number of iterations.
\end{abstract}

\section{Introduction}

We provide some context.

The problem typically addressed by a quantum algorithm can be seen as a
competition between two players. There is a set of functions known to both
players, for example the set of the "periodic" functions $f_{\mathbf{k}%
}\left( x\right) :\left\{ 0,1\right\} ^{n}\rightarrow \left\{ 0,1\right\}
^{n-1}$ -- section 4.1. The first player (the oracle) chooses one of these
functions and gives to the second player a black box (i. e. non-inspectable
inside) hardwired for the computation of that function. The second player
should find out a certain character of the function, for example its period,
by computing $f_{\mathbf{k}}\left( x\right) $ for different values of $x$.
As well known, the quantum algorithm requires a substantially lower number
of function evaluations than the corresponding classical algorithm.

In (Castagnoli, 2008 and 2009), on the basis of theoretical considerations,
we showed that the quantum algorithm requires the number of function
evaluations of a classical algorithm that knows in advance 50\% of the
information that specifies the solution of the problem (see also Castagnoli
and Finkelstein, 2001).

To see this, the key step is representing physically the interdependence
between the problem and the solution: it suffices to represent together the
production of the problem on the part of the oracle and the production of
the solution on the part of the quantum algorithm. We review this step,
which is common to the present work:

(i) We represent the function $f_{\mathbf{k}}\left( x\right) $ chosen by the
oracle by means of an auxiliary quantum register $K$ -- the \textit{oracle's
choice register}. This register, just a conceptual reference, hosts the
suffix $\mathbf{k}$ of $f_{\mathbf{k}}\left( x\right) $, a bit string
defined as follows. In the case of the structured problems, $\mathbf{k}$
represents the table of $f_{\mathbf{k}}\left( x\right) $; in the example of
the "periodic" functions, $\mathbf{k}$ is the sequence of $2^{n}$\ fields of 
$n-1$ bits, the values of the function for increasing values of the
argument. In unstructured data base search, $\mathbf{k}$ is the data base
location chosen by the oracle -- here $f_{\mathbf{k}}\left( x\right) $ is
the Kronecker function $\delta \left( \mathbf{k},x\right) $.

(ii) The black box hardwired for the computation of $f_{\mathbf{k}}\left(
x\right) $ is replaced by a general purpose black box that, given the inputs 
$\mathbf{k}$ (representing the function chosen by the oracle)\ and $x$ (the
argument to query the black box with) computes $f\left( \mathbf{k},x\right)
= $ $f_{\mathbf{k}}\left( x\right) $.

(iii) Register $K$ is ideally added to the usual input register $X$,
containing the value of $x$, and output register $V$, hosting the result of
function evaluation $f\left( \mathbf{k},x\right) $. We should think of
preparing $K$ in an even weighted (indifferently coherent or incoherent)
superposition of all the possible values of $\mathbf{k}$. As usual, $X$ is
prepared in the coherent even weighted superposition of all the possible
values of $x$ and $V$ in a coherent initial state depending on the algorithm.

(iv) Each function evaluation entangles the oracle's choice register $K$ and
the query register $X$. Correspondingly, information about the oracle's
choice propagates from the former to the latter register.

(v) After each function evaluation, we rotate the basis of $X$\ to make this
information readable. Function evaluation/rotation of the $X$\ basis is done
once in the algorithms of Deutsch, Deutsch\&Jozsa, and Simon, iteratively
(for a number of iterations $\func{O}\left( 2^{n/2}\right) $) in Grover's
algorithm.

(vi) Measuring the content of $K$ and $X$ at the end of the algorithm
induces state reduction (i. e. projects the state before measurement) on
both the function chosen by the oracle (the value of $\mathbf{k}$\ hosted in
register $K$) and the solution produced by the algorithm (the value of $x$\
hosted in register $X$)\footnote{%
The state reduction we are dealing with takes nothing from the unitary
character of some quantum algorithms. When the algorithm is unitary, this
reduction entirely originates from the oracle's choice of a value of $%
\mathbf{k}$ out of all possible $\mathbf{k}^{\prime }s$.}. Backdating, to
before running the algorithm, the reduction induced by measuring the content
of $K$\ yields the usual quantum algorithm. \ 

(vii) In this picture, quantum computation is reduction on the solution of
the problem under a relation (or correlation) representing problem-solution
interdependence -- the correlation is between the content of register $K$
(the oracle's choice) and the content of register $X$ (the solution) at the
end of the algorithm.

(viii) Backdating in time, to before running the algorithm, a time symmetric
part of this reduction shows that the quantum algorithm requires the number
of function evaluations of a classical algorithm that knows in advance 50\%
of the information that specifies the solution of the problem (Castagnoli,
2009). We call this the \textit{50\% rule}.

(ix) In this context, the information that specifies the solution of the
problem is the information that specifies both the solution (the content of
register $X$ at the end of the algorithm) and the problem (the content of
register $K$). Since the solution is determined by the content of $K$, the
information that specifies the solution is redundant. Knowing in advance
50\% of the information that specifies the solution of the problem amounts
to knowing in advance 50\% of the information about the solution of the
problem contained in the bit string $\mathbf{k}$ (read in register $K$\ at
the end of the algorithm). The 50\% rule is reformulated as follows: the
quantum algorithm requires the number of function evaluations of a classical
algorithm that knows in advance 50\% of the information about the solution
of the problem contained in $\mathbf{k}$.

(x) In unstructured database search and in Deutsch's problem, $\mathbf{k}$\
is an unstructured bit string. 50\% of the information about the solution of
the problem contained in $\mathbf{k}$\ is represented by any 50\% of the
bits of $\mathbf{k}$ (either one of the two rows of the table of $f_{\mathbf{%
k}}\left( x\right) $ in Deutsch's problem). If instead $\mathbf{k}$ is
structured, which is the case of the structured problems, ascertaining what
is 50\% of the information that specifies the solution of the problem
requires a case by case analysis. It turns out that this information is
represented by those half tables of $f_{\mathbf{k}}\left( x\right) $ that do
not already specify the solution of the problem (the half tables that
specify the solution, containing 100\% of the information about the
solution, should be discarded).

The main objective of this work is checking that the 50\% rule holds for the
main quantum algorithms, namely the algorithms of Deutsch, Deutsch\&Jozsa,
Simon (where the analysis extends by similarity to the hidden subgroup
algorithms, thus in particular to the quantum part of Shor's algorithm), and
Grover.

In Deutsch's and the structured problems, the classical algorithm knowing in
advance 50\% of the rows of the table of $f_{\mathbf{k}}\left( x\right) $
(excluding the half tables that identify the solution) in order to identify
the solution should compute any one\ of the missing rows -- i. e. perform
one function evaluation for any value of $x$ outside the advanced
information. In the structured problems the speed up is exponential since a
classical algorithm without advanced information should compute an
exponential number of rows. In unstructured data base search, knowing in
advance 50\% of the information that specifies the solution of the problem
means knowing in advance 50\% of the $n$\ bits of the data base location
chosen by the oracle. To identify the $n/2$ missing bits, the advanced
information classical algorithm should perform $O\left( 2^{n/2}\right) $
function evaluations. The speed up is quadratic since a classical algorithm
without advanced information should perform $O\left( 2^{n}\right) $ function
evaluations.

Thus the speed up comes from comparing two classical algorithms, with and
without the advanced information. The 50\% rule brings the identification of
the problems solvable with a quantum speed up to the classical framework.

We also show that the advanced information classical algorithm defines the
quantum algorithm. We consider the "skeleton" of the classical algorithm.
This, given the advanced information, performs the function evaluations
required to define the solution of the problem (in the classical framework,
defining does not mean computing). We consider all the possible histories of
this classical algorithm. Each history -- corresponding to a possible way of
getting the advanced information and a possible result of computing the
missing information -- is represented in quantum notation as a sequence of
sharp states. The sum of the histories yields the function evaluation stage
of the quantum algorithm.

The initial phase of each history is chosen in such a way that the transfer
of information from the classical to the quantum algorithm is maximized.

As already said, each function evaluation entangles the oracle's choice
register $K$ and the solution register $X$, information about the oracle's
choice propagates from the former to the latter register.

Which is this information is clear when the entanglement produced by
function evaluation is maximal, like in the algorithms of Deutsch,
Deutsch\&Jozsa (where the only function evaluation produces a maximally
entangled state), and Grover (where each function evaluation feeds the
amplitude of a maximally entangled state at the expense of the amplitude of
an unentangled state). In the maximally entangled state, each orthogonal
state of $K$\ (possibly itself a quantum superposition) corresponds to a
solution of the problem and is correlated with an orthogonal state of $X$.
This allows to define in a constructive way the rotation of the $X$ basis
that makes this solution readable. This completes the definition of the
quantum algorithm (of the algorithm's iterate in Grover's case).

Which is the information propagated to register $X$ is less clear in the
case of Simon's and the hidden subgroup algorithms, where the entanglement
produced by function evaluation is not maximal. The final rotation of the $X$%
\ basis (here the Hadamard transform on $X$) can still be defined as the
transformation that maximizes the information about the oracle's choice
readable in it; this completes the definition of the quantum algorithm,
although it is no more a constructive definition.

Summing up: (i) the 50\% rule brings the search of the speed ups to the
classical framework and (ii) once identified a problem solvable with a
quantum speed up, the same rule could be used a second time for searching
the quantum algorithm that yields the speed up -- in fact the advanced
information classical algorithm defines this quantum algorithm.

In the following sections, for each quantum algorithm: (i) we\ extend the
algorithm to the physical representation of the oracle's choice, (ii) we
check the 50\% rule, and (iii) we rebuild the quantum algorithm out of the
advanced information classical algorithm.

\section{Deutsch's algorithm}

\subsection{Reviewing and extending the algorithm}

We review Deutsch's algorithm (Deutsch, 1985) as revised by (Cleve et al.,
1996). The problem is as follows. An oracle chooses at random one of the
four binary functions $f_{\mathbf{k}}:\left\{ 0,1\right\} \rightarrow
\left\{ 0,1\right\} $, see table (\ref{table}).

\begin{equation}
\begin{tabular}{|c|c|c|c|c|}
\hline
$x$ & $f_{00}(x)$ & $f_{01}(x)$ & $f_{10}(x)$ & $f_{11}(x)$ \\ \hline
0 & 0 & 0 & 1 & 1 \\ \hline
1 & 0 & 1 & 0 & 1 \\ \hline
\end{tabular}%
.  \label{table}
\end{equation}%
$\mathbf{k}\equiv k_{0},k_{1}$, a two-bit string belonging to $\left\{
0,1\right\} ^{2}$; it is both the suffix of $f$\ and, clockwise rotated, the
table of function values ordered for increasing values of the argument -- in
fact $k_{0}=f_{\mathbf{k}}(0)$ and $k_{1}=f_{\mathbf{k}}(1)$. Then the
oracle gives to the second player a black box that, given an input $x$,
computes $f_{\mathbf{k}}(x)$. The second player, by trying function
evaluation for different values of $x$, should find whether the function is
balanced (i. e. $f_{01}$ or $f_{10}$, with an even number of zeroes and
ones) or constant (i. e. $f_{00}$ or $f_{11}$). This requires two function
evaluations in the classical case (for $x=0$\ and $x=1$), just one in the
quantum case. Deutsch's algorithm is the root of all subsequent quantum
algorithms for what concerns both the speed up and the representation of
quantum computation. Bits are replaced by qubits (Finkelstein, 1969,
Benioff, 1982) and reversible logical operations (Bennett, 1973 and 1982,
Fredkin and Toffoli, 1982) by unitary transformations (Finkelstein, 1969,
Benioff, 1982).

We give directly the extension of Deutsch's algorithm to the representation
of the random choice of the function on the part of the oracle (Castagnoli,
2008). A two qubit input register $K$\ contains the oracle's choice $\mathbf{%
k}$. As usual, the one qubit input register $X$ contains the value of $x$ to
query the black box with; the one qubit output register $V$, initially
containing $v$, eventually contains $v\oplus f\left( \mathbf{k},x\right) $
-- the result of function evaluation is module 2 added to $v$\ for logical
reversibility. The black box that, given the input $x$, computes \ $f_{%
\mathbf{k}}(x)$ is replaced by a black box that, given the inputs $\mathbf{k}
$ and $x$, computes $f\left( \mathbf{k},x\right) =f_{\mathbf{k}}(x)$. The
quantum algorithm consists of three steps: (0) prepare register $K$ in an
even weighted superposition of all the possible values of $\mathbf{k}$,
register $X$ in the even weighted superposition of all the possible values
of $x$, and register $V$ in the antisymmetric state, (1)\ perform function
evaluation (which leaves the content of $K$\ and $X$\ unaltered and changes
that of $V$), and (2) apply the Hadamard transform to register $X$.

The initial state is thus%
\begin{equation}
\Psi _{0}=\frac{1}{4}\left( \left\vert 00\right\rangle _{K}+\left\vert
01\right\rangle _{K}+\left\vert 10\right\rangle _{K}+\left\vert
11\right\rangle _{K}\right) \left( \left\vert 0\right\rangle _{X}+\left\vert
1\right\rangle _{X}\right) \left( \left\vert 0\right\rangle _{V}-\left\vert
1\right\rangle _{V}\right) .  \label{initial}
\end{equation}%
The superposition hosted in register $K$\ is indifferently coherent or
incoherent (in this latter case $\left\vert 00\right\rangle _{K}$ should be
replaced by $\func{e}^{i\delta _{00}}\left\vert 00\right\rangle _{K}$, with $%
\delta _{00}$\ a random phase with uniform distribution in $\left[ 0,2\pi %
\right] $, etc.). Function evaluation\ yields

\begin{equation}
\Psi _{1}=\frac{1}{4}\left[ (\left\vert 00\right\rangle _{K}-\left\vert
11\right\rangle _{K})(\left\vert 0\right\rangle _{X}+\left\vert
1\right\rangle _{X})+(\left\vert 01\right\rangle _{K}-\left\vert
10\right\rangle _{K})(\left\vert 0\right\rangle _{X}-\left\vert
1\right\rangle _{X})\right] \left( \left\vert 0\right\rangle _{V}-\left\vert
1\right\rangle _{V}\right) .  \label{fin}
\end{equation}

Applying the Hadamard transform to register $X$\ yields

\begin{equation}
\Psi _{2}=\frac{1}{2\sqrt{2}}\left[ (\left\vert 00\right\rangle
_{K}-\left\vert 11\right\rangle _{K})\left\vert 0\right\rangle
_{X}+(\left\vert 01\right\rangle _{K}-\left\vert 10\right\rangle
_{K})\left\vert 1\right\rangle _{X}\right] \left( \left\vert 0\right\rangle
_{V}-\left\vert 1\right\rangle _{V}\right) .  \label{hadamard}
\end{equation}

Let us denote by $\left[ K\right] $ the content of register $K$, by $\left[ X%
\right] $\ the content of $X$. Measuring $\left[ K\right] $ and $\left[ X%
\right] $\ in (\ref{hadamard}) determines the moves of both players: the
oracle's choice $\mathbf{k}$ in register $K$ and the solution found by the
second player in register $X$. Backdating to before running the algorithm
the state reduction induced by measuring $\left[ K\right] $\ gives the
original Deutsch's algorithm -- it generates at random the value of $\mathbf{%
k}$ hosted in the black box.

\subsection{Checking the 50\% rule}

The information acquired by measuring $\left[ K\right] $ and $\left[ X\right]
$ in (\ref{hadamard})\ is the two bits of the unstructured bit string $%
\mathbf{k}$ -- since the content of $X$ is a function of the content of $K$
the information contained in $X$ is redundant. The quantum algorithm
requires the number of function evaluations of a classical algorithm working
on a solution space reduced in size because one bit of $\mathbf{k}$, either $%
k_{0}=f(\mathbf{k},0)$ or $k_{1}=f(\mathbf{k},1)$, is known in advance. To
identify the character of the function, this algorithm must acquire the
other bit of information by computing, respectively, either $k_{1}=f(\mathbf{%
k},1)$ or $k_{0}=f(\mathbf{k},0)$. Thus the classical algorithm has to
perform one function evaluation like the quantum algorithm.

This verifies the 50\% rule for Deutsch's algorithm. This rule shows that
Deutsch's problem is solvable with a quantum speed up independently of our
knowledge of the quantum algorithm. In fact the speed up comes from
comparing two classical algorithms, with and without the advanced
information.

\subsection{Building the quantum algorithm out of the advanced information
classical algorithm}

We build the function evaluation stage of the quantum algorithm out of the
corresponding stage of a classical algorithm that knows in advance 50\% of $%
\mathbf{k}$. We should consider all the possible ways of getting the
advanced\ information and all the possible results of computing the missing
information, see table (\ref{deutsch}).

\begin{equation}
\begin{tabular}{|c|c|c|c|}
\hline
\# & advanced information & result of function evaluation & character of the
function \\ \hline
1 & $k_{0}=0$ & $k_{1}=f(\mathbf{k},1)=0$ & constant \\ \hline
2 & $k_{0}=0$ & $k_{1}=f(\mathbf{k},1)=1$ & balanced \\ \hline
3 & $k_{0}=1$ & $k_{1}=f(\mathbf{k},1)=0$ & balanced \\ \hline
4 & $k_{0}=1$ & $k_{1}=f(\mathbf{k},1)=1$ & constant \\ \hline
5 & $k_{1}=0$ & $k_{0}=f(\mathbf{k},0)=0$ & constant \\ \hline
6 & $k_{1}=0$ & $k_{0}=f(\mathbf{k},0)=1$ & balanced \\ \hline
7 & $k_{1}=1$ & $k_{0}=f(\mathbf{k},0)=0$ & balanced \\ \hline
8 & $k_{1}=1$ & $k_{0}=f(\mathbf{k},0)=1$ & constant \\ \hline
\end{tabular}
\label{deutsch}
\end{equation}

We represent the possible histories in quantum notation. Since we are
dealing with classical computations, we require that the input and the
output of each history (before and after function evaluation) is a sharp
quantum state. There are sixteen possible histories:

\begin{itemize}
\item Row \#1. The advanced information is $k_{0}=0$. The classical
algorithm should compute $k_{1}=f(\mathbf{k},1)$ that, for this row, is $%
k_{1}=f(\mathbf{k},1)=0$. The quantum representation of the oracle's choice
is thus $\left\vert 00\right\rangle _{K}$. The initial state of register $X$
should be $\left\vert 1\right\rangle _{X}$, the state to query the black box
with in order to compute $f(\mathbf{k},1)$. Since the result of this
computation is module $2$ added to the initial content of register $V$, we
should split row \#1 into two sub-rows: \#1.1 with register $V$ initially in 
$\left\vert 0\right\rangle _{V}$ and \#1.2 with register $V$ initially in $%
\left\vert 1\right\rangle _{V}$. The initial state of history \#1.1 is thus $%
\Psi _{0}^{(1.1)}=\left\vert 00\right\rangle _{K}\left\vert 1\right\rangle
_{X}\left\vert 0\right\rangle _{V}$, that of history \#1.2 is $\Psi
_{0}^{(1.2)}=-\left\vert 00\right\rangle _{K}\left\vert 1\right\rangle
_{X}\left\vert 1\right\rangle _{V}$. These computation histories have to be
added together and must be given an initial phase. For the time being, we
set the initial phase of each history in such a way that, in the
superposition of all histories, we obtain the initial state of the quantum
algorithm; further below we justify this choice independently of our a
priori knowledge of the quantum algorithm.\ To simplify the notation, we sum
together the initial states of these two histories: $\Psi _{0}^{(1)}=\Psi
_{0}^{(1.1)}+\Psi _{0}^{(1.2)}=\left\vert 00\right\rangle _{K}\left\vert
1\right\rangle _{X}\left( \left\vert 0\right\rangle _{V}-\left\vert
1\right\rangle _{V}\right) $. We take care of normalization at the end.
Function evaluation transforms $\Psi _{0}^{(1)}$ into itself: $\Psi
_{1}^{\left( 1\right) }=\Psi _{0}^{(1)}$ (module 2 adding $f\left(
00,1\right) =0$ to the former content of $V$ leaves this content unaltered).

\item Row \#5. Advanced information $k_{1}=0$, result of function evaluation 
$k_{0}=f(\mathbf{k},0)=0$. Applying the same rationale of the above point,
we obtain the initial state $\Psi _{0}^{(5)}=\left\vert 00\right\rangle
_{K}\left\vert 0\right\rangle _{X}\left( \left\vert 0\right\rangle
_{V}-\left\vert 1\right\rangle _{V}\right) $; function evaluation transforms
this state into itself: $\Psi _{1}^{\left( 5\right) }=\Psi _{0}^{(5)}$.

\item The sum of the histories of rows \#1 and \#5 yields the transformation
of $\left\vert 00\right\rangle _{K}\left( \left\vert 0\right\rangle
_{X}+\left\vert 1\right\rangle _{X}\right) \left( \left\vert 0\right\rangle
_{V}-\left\vert 1\right\rangle _{V}\right) $ into itself, namely the
function evaluation stage of Deutsch's algorithm when $K$ is in $\left\vert
00\right\rangle _{K}$.

\item Row \#2. Advanced information $k_{0}=0$; result of function evaluation 
$k_{1}=f(\mathbf{k},1)=1$; initial state $\Psi _{0}^{(2)}=\left\vert
01\right\rangle _{K}\left\vert 1\right\rangle _{X}\left( \left\vert
0\right\rangle _{V}-\left\vert 1\right\rangle _{V}\right) $; state after
function evaluation $\Psi _{1}^{(2)}=-\Psi _{0}^{(2)}$ (module 2 adding $%
f\left( 01,1\right) =1$ to the former content of $V$ swaps $\left\vert
0\right\rangle _{V}$ and $\left\vert 1\right\rangle _{V}$; the overall
result is rotating the phase of the present pair of histories by $\pi $).

\item Row \#7. Advanced information $k_{1}=1$, result of function evaluation 
$k_{0}=f(\mathbf{k},0)=0$; initial state $\Psi _{0}^{(7)}=\left\vert
01\right\rangle _{K}\left\vert 0\right\rangle _{X}\left( \left\vert
0\right\rangle _{V}-\left\vert 1\right\rangle _{V}\right) $; state after
function evaluation $\Psi _{1}^{(7)}=\Psi _{0}^{(7)}$(module 2 adding $%
f\left( 01,0\right) =0$ to the former content of $V$ leaves this content
unaltered).

\item The sum of the histories of rows \#2 and \#7 yields the transformation
of $\left\vert 01\right\rangle _{K}\left( \left\vert 0\right\rangle
_{X}+\left\vert 1\right\rangle _{X}\right) \left( \left\vert 0\right\rangle
_{V}-\left\vert 1\right\rangle _{V}\right) $ into $\left\vert
01\right\rangle _{K}\left( \left\vert 0\right\rangle _{X}-\left\vert
1\right\rangle _{X}\right) \left( \left\vert 0\right\rangle _{V}-\left\vert
1\right\rangle _{V}\right) $, namely the function evaluation stage of
Deutsch's algorithm when $K$ is in $\left\vert 01\right\rangle _{K}$.

\item We proceed in a similar way for the other histories. Summing together
the quantum representations of the $16$\ classical histories and normalizing
yields the transformation of $\Psi _{0}$ (equation \ref{initial}) into $\Psi
_{1}$ (equation \ref{fin}).
\end{itemize}

In hindsight, we can see a shortcut. For each $\left\vert \mathbf{k}%
\right\rangle _{K}$,\ we perform function evaluation not only for those
values of $x$ required to identify the solution of the problem, but also for
all the other possible values of $x$. In other words, we perform function
evaluation for each product $\left\vert \mathbf{k}\right\rangle _{K}\left(
\left\vert 0\right\rangle _{X}+\left\vert 1\right\rangle _{X}\right) \left(
\left\vert 0\right\rangle _{V}-\left\vert 1\right\rangle _{V}\right) $;\
junk histories (for that $\left\vert \mathbf{k}\right\rangle _{K}$) do not
harm, the important thing is performing function evaluation for the values
of $x$ required to identify the solution. As one can see, this yields
directly the transformation of $\Psi _{0}$ (equation \ref{initial}) into $%
\Psi _{1}$ (equation \ref{fin}). Conversely, by simply inspecting the form
of $\Psi _{0}$ in equation (\ref{initial}), one can see that each $%
\left\vert \mathbf{k}\right\rangle _{K}\left( \left\vert 0\right\rangle
_{X}+\left\vert 1\right\rangle _{X}\right) \left( \left\vert 0\right\rangle
_{V}-\left\vert 1\right\rangle _{V}\right) $ is the initial state of a bunch
of histories as from the above shortcut.

Summing up, quantum parallel computation can be seen as the sum of the
histories of a classical algorithm that, given the advanced information,
computes the missing information required to identify the solution of the
problem. This holds in general for the function evaluation stage of all
quantum algorithms.

By considering the sum of the histories, we can justify the choice of the
initial phase. We take the generic initial state of register $V$: $\alpha
\left( \left\vert 0\right\rangle _{V}+\left\vert 1\right\rangle _{V}\right)
+\beta \left( \left\vert 0\right\rangle _{V}-\left\vert 1\right\rangle
_{V}\right) $ -- the initial phase of the histories with register $V$ in $%
\left\vert 0\right\rangle _{V}$ is $\alpha +\beta $, that of the histories
with $V$ in $\left\vert 1\right\rangle _{V}$ is $\alpha -\beta $. Under the
amplitude $\alpha $, the computation performed by the advanced information
classical algorithm gets lost in the quantum translation, since the overall
initial state is transformed into itself. Under $\beta $, the transfer of
information from the classical to the quantum algorithm is maximum (we
obtain the above development). This justifies the choice $\alpha =0$ and $%
\beta =1$.

Now we look at the outcome of the second stage -- equation (\ref{fin}).
Function evaluation has created a maximal entanglement between registers $K$
and $X$, two orthogonal states of $K$, $\left\vert 00\right\rangle
_{K}-\left\vert 11\right\rangle _{K}$ and $\left\vert 01\right\rangle
_{K}-\left\vert 10\right\rangle _{K}$ (or indifferently $\func{e}^{i\delta
_{00}}\left\vert 00\right\rangle _{K}-\func{e}^{i\delta _{11}}\left\vert
11\right\rangle _{K}$ and $\func{e}^{i\delta _{01}}\left\vert
01\right\rangle _{K}-\func{e}^{i\delta _{10}}\left\vert 10\right\rangle _{K}$%
) are correlated with two orthogonal states of $X$, respectively $\left\vert
0\right\rangle _{X}+\left\vert 1\right\rangle _{X}$ and $\left\vert
0\right\rangle _{X}-\left\vert 1\right\rangle _{X}$. This means that, after
function evaluation, register $X$\ contains the information that
discriminates between $\left\vert 00\right\rangle _{K}-\left\vert
11\right\rangle _{K}$ and $\left\vert 01\right\rangle _{K}-\left\vert
10\right\rangle _{K}$, namely between constant and balanced functions.
Therefore we should rotate the $X$\ basis in such a way that this
information becomes readable: $\left\vert 0\right\rangle _{X}+\left\vert
1\right\rangle _{X}$\ should be transformed into $\left\vert 0\right\rangle
_{X}$, etc.. This is a constructive definition of the Hadamard transform on
register $X$. This completes the derivation of the quantum algorithm from
the classical algorithm with the advanced information.

\section{Deutsch\&Jozsa algorithm}

\subsection{Reviewing and extending the algorithm}

Deutsch\&Jozsa's algorithm is a generalization of Deutsch's algorithm that
achieves an exponential speed up (Deutsch and Jozsa, 1989). This time we
deal with the set of the functions $f_{\mathbf{k}}:\left\{ 0,1\right\}
^{n}\rightarrow \left\{ 0,1\right\} $ such that the function is either
constant (all zeroes or all ones), or balanced (even number of zeroes and
ones). $\mathbf{k}\equiv k_{0},k_{1},...,k_{2^{n}-1}$ is a $2^{n}$ bit
string rpresenting the table of the function -- namely the sequence of
function values for increasing values of the argument. Table (\ref{dj})
shows this set of functions for $n=2$ -- we shall focus on this example. 
\begin{equation}
\begin{tabular}{|c|c|c|c|c|c|c|c|c|}
\hline
$x$ & $\,f_{0000}\left( x\right) $ & $f_{1111}\left( x\right) $ & $%
f_{0011}\left( x\right) $ & $f_{1100}\left( x\right) $ & $f_{0101}\left(
x\right) $ & $f_{1010}\left( x\right) $ & $f_{0110}\left( x\right) $ & $%
f_{1001}\left( x\right) $ \\ \hline
00 & 0 & 1 & 0 & 1 & 0 & 1 & 0 & 1 \\ \hline
01 & 0 & 1 & 0 & 1 & 1 & 0 & 1 & 0 \\ \hline
10 & 0 & 1 & 1 & 0 & 0 & 1 & 1 & 0 \\ \hline
11 & 0 & 1 & 1 & 0 & 1 & 0 & 0 & 1 \\ \hline
\end{tabular}
\label{dj}
\end{equation}%
Note that $k_{0}=f_{\mathbf{k}}(00),$ $k_{1}=f_{\mathbf{k}}(01),$ $k_{2}=f_{%
\mathbf{k}}(10)$ and $k_{3}=f_{\mathbf{k}}(11)$. The string $\mathbf{k}$\ is
both the suffix of $f$\ and, clockwise rotated, the table of the function
chosen by the oracle.

The problem is as follows. An oracle chooses at random one of these
functions and gives to the second player the black box hardwired for the
computation of that function. The second player, by trying function
evaluation for different values of $x$, must find whether the function is
balanced or constant. In the worst case, this requires a number of function
evaluations $\exp \left( n\right) $\ in the classical case, just one in the
quantum case.

We give directly the extension of the quantum algorithm to the
representation of the choice of the function on the part of the oracle. The
black box, given the inputs $\mathbf{k}$ and $x$, computes $f\left( \mathbf{k%
},x\right) =f_{\mathbf{k}}(x)$. The $2^{n}$ qubit oracle's choice \ register 
$K$\ contains the table of the function -- we should keep in mind that this
register is just a conceptual reference. The query register $X$ is $n$
qubit. The output register $V$ is one qubit. The algorithm consists of three
steps: (0) prepare register $K$ in an even weighted superposition of all the
possible values of $\mathbf{k}$, register $X$ in the even weighted
superposition of all the possible values of $x$, and register $V$ in the
antisymmetric state, (1)\ perform function evaluation, which changes the
content of $V$ from $v$ to $v\oplus f\left( \mathbf{k},x\right) $, and (2)
apply the Hadamard transform to register $X$.

The initial state is

\begin{eqnarray}
\Psi _{0} &=&\frac{1}{8}\left( \left\vert 0000\right\rangle _{K}+\left\vert
1111\right\rangle _{K}+\left\vert 0011\right\rangle _{K}+\left\vert
1100\right\rangle _{K}+...\right)  \label{indj} \\
&&\left( \left\vert 00\right\rangle _{X}+\left\vert 01\right\rangle
_{X}+\left\vert 10\right\rangle _{X}+\left\vert 11\right\rangle _{X}\right)
\left( \left\vert 0\right\rangle _{V}-\left\vert 1\right\rangle _{V}\right) .
\nonumber
\end{eqnarray}

Function evaluation yields

\begin{equation}
\Psi _{1}=\frac{1}{4}\left[ 
\begin{array}{c}
(\left\vert 0000\right\rangle _{K}-\left\vert 1111\right\rangle
_{K})(\left\vert 00\right\rangle _{X}+\left\vert 01\right\rangle
_{X}+\left\vert 10\right\rangle _{X}+\left\vert 11\right\rangle _{X})+ \\ 
(\left\vert 0011\right\rangle _{K}-\left\vert 1100\right\rangle
_{K})(\left\vert 00\right\rangle _{X}+\left\vert 01\right\rangle
_{X}-\left\vert 10\right\rangle _{X}-\left\vert 11\right\rangle _{X})+...%
\end{array}%
\right] \left( \left\vert 0\right\rangle _{V}-\left\vert 1\right\rangle
_{V}\right) .  \label{evaluation}
\end{equation}%
Applying the Hadamard transform to register $X$\ yields

\begin{equation}
\Psi _{2}=\frac{1}{4}\left[ \left( \left\vert 0000\right\rangle
_{K}-\left\vert 1111\right\rangle _{K}\right) \left\vert 00\right\rangle
_{X}+(\left\vert 0011\right\rangle _{K}-\left\vert 1100\right\rangle
_{K})\left\vert 10\right\rangle _{X}+....\right] \left( \left\vert
0\right\rangle _{V}-\left\vert 1\right\rangle _{V}\right) .  \label{hdj}
\end{equation}%
Measuring $\left[ K\right] $\ and $\left[ X\right] $\ in (\ref{hdj})
determines the moves of both players, namely the oracle's choice (a value of 
$\mathbf{k}$) and the solution provided by the second player: all zeroes if
the function is constant, not so if the function is balanced. Backdating to
before running the algorithm the state reduction induced by measuring $\left[
K\right] $\ gives the original Deutsch\&Jozsa's algorithm -- it generates at
random the value of $\mathbf{k}$ hosted in the black box.

\subsection{Checking the 50\% rule}

The information acquired by measuring $\left[ K\right] $ and $\left[ X\right]
$ in (\ref{hdj})\ is the information in the structured bit string $\mathbf{k}
$ -- the table of the function $f_{\mathbf{k}}\left( x\right) $. Since the
content of $X$ is a function of the content of $K$, the information
contained in $X$ is redundant. 50\% of the information about the solution
(in $\mathbf{k}$) is represented by all the possible half tables that do not
contain different values of the function. Those with different values
already say that the function is balanced -- they provide 100\% of the
information about the solution and should be discarded. For example, with
reference to table (\ref{dj}), let us consider the half tables for $x=00$
and $x=01$. Those in the first four columns contain exactly 50\% of the
information in $\mathbf{k}$ (which is also evident for reasons of symmetry)\
and thus represent the advanced information. The half tables in the last
four columns (for $x=00$ and $x=01$) already say that the function is
balanced and are discarded. By the way, this does not mean discarding these
columns; e. g. in the fifth column, the half table for $x=00$ and $x=10$ and
that for $x=01$ and $x=11$\ are good.

With this definition of the advanced information, the solution is always
identified by computing an extra row, namely by performing one function
evaluation for any value of $x$\ outside the advanced information (if the
value of the function is still the same, the function is constant, otherwise
it is balanced).

This verifies the 50\% rule for Deutsch\&Jozsa algorithm. This rule shows
that Deutsch\&Jozsa's problem is solvable with an exponential speed up
independently of our knowledge of the quantum algorithm -- the speed up
comes from comparing two classical algorithms, with and without the advanced
information.

\subsection{Building the quantum algorithm out of the advanced information
classical algorithm}

The function evaluation stage of the quantum algorithm -- namely the
transformation of $\Psi _{0}$ (equation \ref{indj}) into $\Psi _{1}$
(equation \ref{evaluation}) -- is the sum of the histories of the advanced
information classical algorithm. This is clear by looking at the form of $%
\Psi _{0}$ with the shortcut of section 2.3 in mind. Without shortcut we
obtain the same result, as follows:

\begin{itemize}
\item Register $K$ in $\left\vert 0000\right\rangle _{K}$; advanced
information $k_{0}=0$ and $k_{1}=0$; in order to ascertain whether the
function is constant or balanced, we should perform function evaluation for
either $x=10$ or $x=11$;\ as we are building the superposition of all the
possible histories, we do it for the superposition of $x=10$ and $x=11$;
with $K$ in $\left\vert 0000\right\rangle _{K}$, the result of function
evaluation is $k_{2}=0$ and $k_{3}=0$; thus the initial state is $\Psi
_{0}^{(1)}=\left\vert 0000\right\rangle _{K}\left( \left\vert
10\right\rangle _{X}+\left\vert 11\right\rangle _{X}\right) \left(
\left\vert 0\right\rangle _{V}-\left\vert 1\right\rangle _{V}\right) $ and
the outcome of function evaluation is $\Psi _{1}^{\left( 1\right) }=\Psi
_{0}^{(1)}$ (module 2 adding $f_{0000}(10)=0$ or $f_{0000}(11)=0$ to the
former content of $V$ leaves this content unaltered).

\item Register $K$ in $\left\vert 0000\right\rangle _{K}$; advanced
information $k_{2}=0$ and $k_{3}=0$; result of function evaluation (for $%
x=10 $ or $x=11$) $k_{0}=0$ and $k_{1}=0$; initial state $\Psi
_{0}^{(2)}=\left\vert 0000\right\rangle _{K}\left( \left\vert
01\right\rangle _{X}+\left\vert 00\right\rangle _{X}\right) \left(
\left\vert 0\right\rangle _{V}-\left\vert 1\right\rangle _{V}\right) $,
outcome of function evaluation $\Psi _{1}^{\left( 2\right) }=\Psi _{0}^{(2)}$%
.

\item The sum of the histories \#1 and \#2 yields the function evaluation
stage of Deutsch\&Jozsa's algorithm when $K$ is in $\left\vert
0000\right\rangle _{K}$; with normalization, considering other ways of
getting the advanced information (with $K$ in $\left\vert 0000\right\rangle
_{K}$) does not modify the result already obtained.

\item Register $K$ in $\left\vert 0101\right\rangle _{K}$; advanced
information, e. g., $k_{0}=0$ and $k_{1}=1$; we already know that the
function is balanced, no function evaluation is needed, and there are no
histories in such a case.
\end{itemize}

By summing together all the histories corresponding to the good half tables,
we obtain the function evaluation stage of the quantum algorithm.

The choice of the initial phase of each history is justified as in Deutsch's
algorithm.

We examine the outcome of function evaluation, namely $\Psi _{1}$ (equation %
\ref{evaluation}). There is a maximal entanglement between registers $K$\
and $X$.\ Orthogonal states of $K$\ , discriminating between constant and
balanced functions (also between different kinds of balanced functions, but
this is not relevant), are correlated with orthogonal states of $X$. This
means that the information whether the function is constant or balanced has
propagated to register $X$. To read this information, we should rotate the $%
X $ basis in such a way that $(\left\vert 0000\right\rangle _{K}-\left\vert
1111\right\rangle _{K})(\left\vert 00\right\rangle _{X}+\left\vert
01\right\rangle _{X}+\left\vert 10\right\rangle _{X}+\left\vert
11\right\rangle _{X}$\ goes into $(\left\vert 0000\right\rangle
_{K}-\left\vert 1111\right\rangle _{K})\left\vert 00\right\rangle _{X}$,
etc.. This is a constructive definition of the Hadamard transform on
register $X$. This completes the derivation of the quantum algorithm from
the classical algorithm with the advanced information.

\section{Simon's algorithm}

\subsection{Reviewing and extending the algorithm}

We deal with the set of the "periodic" functions $f_{\mathbf{k}}:\left\{
0,1\right\} ^{n}\rightarrow \left\{ 0,1\right\} ^{n-1}$. The "periodic"
function $f_{\mathbf{k}}\left( x\right) $ is such that $f_{\mathbf{k}}\left(
x\right) =f_{\mathbf{k}}\left( y\right) $ if and only if $x=y$\ or $%
x=y\oplus \mathbf{h}^{\left( \mathbf{k}\right) }$, where: (i) $\mathbf{k}%
\equiv k_{0},k_{1},...,k_{2^{n}\left( n-1\right) -1}$ is a $2^{n}\left(
n-1\right) $ bit string, the sequence of function values (each a field of $%
n-1$ bits) for increasing values of the argument, (ii) $\mathbf{h}^{\left( 
\mathbf{k}\right) }\mathbf{\equiv ~}h_{0}^{\left( \mathbf{k}\right)
},h_{1}^{\left( \mathbf{k}\right) },...,h_{n-1}^{\left( \mathbf{k}\right) }$
is an $n$ bit string (depending on the value of $\mathbf{k}$) belonging to $%
\left\{ 0,1\right\} ^{n}$ with the exclusion of the all zeroes string, (iii) 
$x$ and $y$ are variables belonging to $\left\{ 0,1\right\} ^{n}$\ also
represented as $n$\ bit strings, and (iv) $\oplus $\ denotes bit by bit
module $2$ addition.

Thus, the string $\mathbf{h}^{\left( \mathbf{k}\right) }$,\ also called the 
\textit{hidden string, }is a sort of period of the function $f_{\mathbf{k}%
}\left( x\right) $. Since $\mathbf{h}^{\left( \mathbf{k}\right) }\oplus 
\mathbf{h}^{\left( \mathbf{k}\right) }=0$, each value of the function
appears exactly twice in the table of the function. This means that 50\% of
the rows plus one surely contain a same value twice, which identifies the
period. \ 

By way of exemplification, table (\ref{periodic}) gives the set of the
periodic functions for $n=2$. 
\begin{equation}
\begin{tabular}{|c|c|c|c|c|c|c|}
\hline
& $\mathbf{h}^{\left( 0011\right) }=01$ & $\mathbf{h}^{\left( 1100\right)
}=01$ & $\mathbf{h}^{\left( 0101\right) }=10$ & $\mathbf{h}^{\left(
1010\right) }=10$ & $\mathbf{h}^{\left( 0110\right) }=11$ & $\mathbf{h}%
^{\left( 1001\right) }=11$ \\ \hline
$x$ & $f_{0011}\left( x\right) $ & $f_{1100}\left( x\right) $ & $%
f_{0101}\left( x\right) $ & $f_{1010}\left( x\right) $ & $f_{0110}\left(
x\right) $ & $f_{1001}\left( x\right) $ \\ \hline
00 & 0 & 1 & 0 & 1 & 0 & 1 \\ \hline
01 & 0 & 1 & 1 & 0 & 1 & 0 \\ \hline
10 & 1 & 0 & 0 & 1 & 1 & 0 \\ \hline
11 & 1 & 0 & 1 & 0 & 0 & 1 \\ \hline
\end{tabular}
\label{periodic}
\end{equation}%
Note that $k_{0}=f_{\mathbf{k}}(00),$ $k_{1}=f_{\mathbf{k}}(01),$ $k_{2}=f_{%
\mathbf{k}}(10),$ and $k_{3}=f_{\mathbf{k}}(11)$: the string $\mathbf{k}$\
is both the suffix of $f$\ and, clockwise rotated, the table of function
values for increasing values of the argument.

The problem is as follows. The oracle chooses at random a function $f_{%
\mathbf{k}}\left( x\right) $, then he gives to the second player the black
box hardwired for the computation of that function. The second player should
find the string $\mathbf{h}^{\left( \mathbf{k}\right) }$\ (the "period" of
the function)\ by performing function evaluation for different values of $x$.

To find $\mathbf{h}^{\left( \mathbf{k}\right) }$ with probability, say, $%
\frac{2}{3}$, $f_{\mathbf{k}}\left( x\right) $ must be computed the order of 
$2^{\frac{n}{3}}$ times in the classical case, the order of $3n$ times in
the quantum case. There is an exponential speed up (Simon, 1994).

We give directly the extension of the quantum algorithm to the
representation of the choice of the function on the part of the oracle. The
black box, given the inputs $\mathbf{k}$ and $x$, computes $f\left( \mathbf{k%
},x\right) =f_{\mathbf{k}}(x)$. The oracle's choice \ register $K$\ is $%
2^{n}\left( n-1\right) $ qubit. The query register $X$ is $n$ qubit. The
output register $V$ is $n-1$ qubit. The algorithm consists of three steps:
(0) prepare register $K$ in an even weighted superposition of all the
possible values of $\mathbf{k}$, register $X$ in the even weighted
superposition of all the possible values of $x$, and register $V$ in the all
zeroes string $\left\vert 0\right\rangle _{V}$, (1)\ perform function
evaluation, which changes the content of $V$ from $\mathbf{v}$ (an $n-1$\
bit string) to $\mathbf{v}\oplus f\left( \mathbf{k},x\right) $, where $%
\oplus $ denotes bit by bit module $2$ addition, and (2) apply the Hadamard
transform to register $X$.

The initial state is

\begin{equation}
\Psi _{0}=\frac{1}{4\sqrt{3}}\left( \left\vert 0011\right\rangle
_{K}+\left\vert 1100\right\rangle _{K}+\left\vert 0101\right\rangle
_{K}+\left\vert 1010\right\rangle _{K}+...\right) \left( \left\vert
00\right\rangle _{X}+\left\vert 01\right\rangle _{X}+\left\vert
10\right\rangle _{X}+\left\vert 11\right\rangle _{X}\right) \left\vert
0\right\rangle _{V}.  \label{insimon}
\end{equation}%
Function evaluation yields

\begin{equation}
\Psi _{1}=\frac{1}{4\sqrt{3}}\left[ 
\begin{array}{c}
(\left\vert 0011\right\rangle _{K}+\left\vert 1100\right\rangle _{K})\left[
(\left\vert 00\right\rangle _{X}+\left\vert 01\right\rangle _{X})\left\vert
0\right\rangle _{V}+(\left\vert 10\right\rangle _{X}+\left\vert
11\right\rangle _{X})\left\vert 1\right\rangle _{V}\right] + \\ 
(\left\vert 0101\right\rangle _{K}+\left\vert 1010\right\rangle _{K})\left[
(\left\vert 00\right\rangle _{X}+\left\vert 10\right\rangle _{X})\left\vert
0\right\rangle _{V}+(\left\vert 01\right\rangle _{X}+\left\vert
11\right\rangle _{X})\left\vert 1\right\rangle _{V}\right] +...%
\end{array}%
\right] .  \label{second}
\end{equation}

Applying the Hadamard transform to register $X$\ yields

\begin{equation}
\Psi _{2}=\frac{1}{4}\left[ 
\begin{array}{c}
(\left\vert 0011\right\rangle _{K}+\left\vert 1100\right\rangle _{K})\left[
(\left\vert 00\right\rangle _{X}+\left\vert 10\right\rangle _{X})\left\vert
0\right\rangle _{V}+(\left\vert 00\right\rangle _{X}-\left\vert
10\right\rangle _{X})\left\vert 1\right\rangle _{V}\right] + \\ 
(\left\vert 0101\right\rangle _{K}+\left\vert 1010\right\rangle _{K})\left[
(\left\vert 00\right\rangle _{X}+\left\vert 01\right\rangle _{X})\left\vert
0\right\rangle _{V}+(\left\vert 00\right\rangle _{X}-\left\vert
01\right\rangle _{X})\left\vert 1\right\rangle _{V}\right] +...%
\end{array}%
\right] .  \label{hsimon}
\end{equation}%
Backdating to before running the algorithm the state reduction induced by
measuring $\left[ K\right] $\ gives the original Simon's algorithm -- it
generates at random the value of $\mathbf{k}$ hosted in the black box.

As one can see (equation \ref{hsimon}), for each pair of complementary
values of the oracle's choice (e. g. for register $K$\ in $\left\vert
0011\right\rangle _{K}+\left\vert 1100\right\rangle _{K}$)\ and for each
value of $f_{\mathbf{k}}\left( x\right) $ (e. g. for register $V$ in $%
\left\vert 0\right\rangle _{V}$), register $X$ hosts an even weighted\
superposition (e. g. $\left\vert 00\right\rangle _{X}+\left\vert
10\right\rangle _{X}$) of the $2^{n-1}$ strings $\mathbf{s}_{j}^{\left( 
\mathbf{k}\right) }$ ($j=0,1,...,2^{n-1}-1$) "orthogonal" to the hidden
string $\mathbf{h}^{\left( \mathbf{k}\right) }$ (if we multiply bit by bit
two orthogonal strings and take the module $2$ addition of the product bits,
the result is zero) -- $00$\ and $10$\ are the two strings orthogonal to the
hidden strings $\mathbf{h}^{\left( 0011\right) }=\mathbf{h}^{\left(
1100\right) }=01$. Note that, in (\ref{hsimon}), only the phase of the terms
of this superposition depend on the value of $f_{\mathbf{k}}\left( x\right) $%
. Therefore, by measuring $\left[ K\right] $\ and $\left[ X\right] $ in (\ref%
{hsimon}), we obtain at random the oracle's choice $\mathbf{k}$ and one of
the $\mathbf{s}_{j}^{\left( \mathbf{k}\right) }$ orthogonal to $\mathbf{h}%
^{\left( \mathbf{k}\right) }$.

At this point, we leave register $K$ in its after-measurement state, so that
the value of $\mathbf{k}$ remains fixed, and iterate the right part of the
algorithm (initial preparation of registers $X$\ and $V$, function
evaluation, Hadamard transform on $X$, and measurement of $\left[ X\right] $%
) until obtaining $n-1$ different $\mathbf{s}_{j}^{\left( \mathbf{k}\right)
} $. This allows to find $\mathbf{h}^{\left( \mathbf{k}\right) }$ by solving
a system of $n-1$ module $2$ linear equations. If the algorithm is iterated,
say, $3n$ times, the probability of obtaining $n-1$ different $\mathbf{h}%
_{j}^{\left( \mathbf{k}\right) }$, thus of finding the solution, is $\frac{2%
}{3}$. The probability of not finding the solution goes down exponentially
with the number of iterations.

\subsection{Checking the 50\% rule}

For the sake of simplicity, we reformulate Simon's problem as the problem of
generating at random a string $\mathbf{s}_{j}^{\left( \mathbf{k}\right) }$
orthogonal to $\mathbf{h}^{\left( \mathbf{k}\right) }$ -- rather than
finding the hidden string $\mathbf{h}^{\left( \mathbf{k}\right) }$. Any $%
\mathbf{s}_{j}^{\left( \mathbf{k}\right) }$ is thus a "solution of the
problem". For what concerns the character of the speed up, the two
formulations are equivalent: an exponential speed up in the former implies
an exponential speed up in the latter and vice-versa.

The information acquired by measuring $\left[ K\right] $ and $\left[ X\right]
$ in (\ref{hsimon})\ is the information in the structured bit string $%
\mathbf{k}$, the table of the function $f_{\mathbf{k}}\left( x\right) $ --
since the content of $X$ (the string $\mathbf{s}_{j}^{\left( \mathbf{k}%
\right) }$) is a function of the content of $K$, the information contained
in $X$ is redundant. As one can check, e. g. in table (\ref{periodic}), 50\%
of the information in $\mathbf{k}$ is represented by the half tables that do
not contain a same value of the function twice (which already identifies the
period and thus any $\mathbf{s}_{j}^{\left( \mathbf{k}\right) }$).

With this definition of the advanced information, the solution is always
identified by computing an extra row, namely by performing one function
evaluation for any value of $x$\ outside the advanced information (because
of the structure of the problem, the new value of the function is
necessarily a value already present in the advanced information).

This verifies the 50\% rule for Simon's algorithm. This rule shows that
Simon's problem is solvable with an exponential speed up independently of
our knowledge of the quantum algorithm -- the speed up comes from comparing
two classical algorithms, with and without the advanced information.

One can readily see that the same holds by similarity for the generalized
Simon's algorithm, thus for the hidden subgroup algorithms (Mosca and Ekert,
1999), like finding orders, finding the period of a function (the quantum
part of Shor's factorization algorithm), finding discrete logarithms, etc.
(e. g., Kaye et al., 2007).

\subsection{Building the quantum algorithm out of the advanced information
classical algorithm}

The function evaluation stage of the quantum algorithm -- namely the
transformation of $\Psi _{0}$ (equation \ref{insimon}) into $\Psi _{1}$
(equation \ref{second}) -- is the sum of the histories of the advanced
information classical algorithm (the rationale is the same of sections 2.3
and 3.3).

We justify the choice of the initial state of $V$. We start with\ the
generic initial state $\alpha \left\vert 0\right\rangle _{V}+\beta
\left\vert 1\right\rangle _{V}$ ($\alpha $ is thus the initial phase of the
histories beginning with $\left\vert 0\right\rangle _{V}$, etc.). Under $%
\alpha $, we obtain the transformation of $\Psi _{0}$\ (equation \ref%
{insimon}) into $\Psi _{1}$ (equation \ref{second}). Under $\beta $, we
obtain the same result with $\left\vert 0\right\rangle _{V}$ and $\left\vert
1\right\rangle _{V}$ interchanged. Since we are interested in the
superposition hosted in register $X$, which is the same in either case, we
can suppress either $\alpha $ or $\beta $.

We examine the outcome of function evaluation, namely $\Psi _{1}$ (equation %
\ref{second}). This time the entanglement between registers $K$ and $X$\ is
not maximal. We know that function evaluation moves to register $X$
information about the oracle's choice, but we do not know which is this
information. The Hadamard transform on register $X$ can still be defined as
the rotation of the $X$\ basis that maximizes the information about the
oracle's choice readable in it. However, this is no more a constructive
definition, we are left with the problem of discovering that this
information is a string orthogonal to $\mathbf{h}^{\left( \mathbf{k}\right)
} $.

\section{Grover's algorithm}

\subsection{Reviewing and extending the algorithm}

The problem addressed by Grover's algorithm (Grover, 1996) is data base
search. We deal with the set of the Kronecker functions $\delta \left( 
\mathbf{k},x\right) $, where $\mathbf{k}\equiv \mathbf{~}%
k_{0},k_{1},...,k_{n-1}$ is an $n$\ bit string belonging to $\left\{
0,1\right\} ^{n}$. The oracle chooses one of these functions -- chooses the
data base location $\mathbf{k}$\ -- and gives to the second player the black
box hardwired for the computation of that function. The second player has to
find the value of $\mathbf{k}$ by computing $\delta \left( \mathbf{k}%
,x\right) $ for different values of $x$. In the classical case $\delta
\left( \mathbf{k},x\right) $\ must be computed $O\left( 2^{n}\right) $\
times, in the quantum case $O\left( 2^{n/2}\right) $\ times. There is a
quadratic speed up.

We give directly the extension of the quantum algorithm to the
representation of the choice of the function on the part of the oracle. The
black box, given the inputs $\mathbf{k}$ and $x$, computes $\delta \left( 
\mathbf{k},x\right) $. The $n$ qubit register $K$ contains the bit string $%
\mathbf{k}$ -- the data base location chosen by the oracle. The query
register $X$ is $n$ qubit. The output register $V$ is one qubit. The
algorithm consists of three steps: (0) prepare register $K$ in an even
weighted superposition of all the possible values of $\mathbf{k}$, register $%
X$ in the even weighted superposition of all the possible values of $x$, and
register $V$ in the antisymmetric state, (1)\ perform function evaluation,
which changes the content of $V$ from $v$ to $v\oplus \delta \left( \mathbf{k%
},x\right) $, and (2) apply the transformation $U$ (see further below) to
register $X$.

We start with $n=2$, then generalize to $n>2$. The initial state is:

\begin{equation}
\text{ }\Psi _{0}=\frac{1}{4\sqrt{2}}\left( \left\vert 00\right\rangle
_{K}+\left\vert 01\right\rangle _{K}+\left\vert 10\right\rangle
_{K}+\left\vert 11\right\rangle _{K}\right) \left( \left\vert
00\right\rangle _{X}+\left\vert 01\right\rangle _{X}+\left\vert
10\right\rangle _{X}+\left\vert 11\right\rangle _{X}\right) (\left\vert
0\right\rangle _{V}-\left\vert 1\right\rangle _{V}).  \label{preparation}
\end{equation}%
Function evaluation yields:

\begin{equation}
\Psi _{1}=\frac{1}{4\sqrt{2}}\left[ 
\begin{array}{c}
\left\vert 00\right\rangle _{K}\left( -\left\vert 00\right\rangle
_{X}+\left\vert 01\right\rangle _{X}+\left\vert 10\right\rangle
_{X}+\left\vert 11\right\rangle _{X}\right) + \\ 
\left\vert 01\right\rangle _{K}\left( \left\vert 00\right\rangle
_{X}-\left\vert 01\right\rangle _{X}+\left\vert 10\right\rangle
_{X}+\left\vert 11\right\rangle _{X}\right) + \\ 
\left\vert 10\right\rangle _{K}\left( \left\vert 00\right\rangle
_{X}+\left\vert 01\right\rangle _{X}-\left\vert 10\right\rangle
_{X}+\left\vert 11\right\rangle _{X}\right) + \\ 
\left\vert 11\right\rangle _{K}\left( \left\vert 00\right\rangle
_{X}+\left\vert 01\right\rangle _{X}+\left\vert 10\right\rangle
_{X}-\left\vert 11\right\rangle _{X}\right)%
\end{array}%
\right] (\left\vert 0\right\rangle _{V}-\left\vert 1\right\rangle _{V}).
\label{secondstage}
\end{equation}%
This is four orthogonal states of $K$\ correlated with four orthogonal
states of $X$. \ Applying to register $X$ the Hadamard transform,\ then the
transformation obtained by computing $\delta \left( 0,x\right) $, then
another time the Hadamard transform (for short, applying the transformation $%
U$) yields:

\begin{equation}
\Psi _{2}=\frac{1}{2\sqrt{2}}\left( \left\vert 00\right\rangle
_{K}\left\vert 00\right\rangle _{X}+\left\vert 01\right\rangle
_{K}\left\vert 01\right\rangle _{X}+\left\vert 10\right\rangle
_{K}\left\vert 10\right\rangle _{X}+\left\vert 11\right\rangle
_{K}\left\vert 11\right\rangle _{X}\right) (\left\vert 0\right\rangle
_{V}-\left\vert 1\right\rangle _{V}),  \label{final}
\end{equation}%
namely an entangled state where each value of $k$ is correlated with the
corresponding solution found by the second player (the same value of $k$ but
in register $X$). The final measurement of $\left[ K\right] $ and $\left[ X%
\right] $ in state (\ref{final})\ determines the moves of both players, the
oracle's choice (the value of $\mathbf{k}$) and the solution provided by the
second player. The reduction induced by measuring $\left[ K\right] $,
backdated to before running the algorithm, yields the original Grover's
algorithm.

\subsection{Checking the 50\% rule}

The information acquired in the final measurement of $\left[ K\right] $\ and 
$\left[ X\right] $ in (\ref{final}) is the two bits of the unstructured bit
string $\mathbf{k}$ -- since the content of $X$\ is a function of the
content of $K$ the information contained in $X$ is redundant. The quantum
algorithm requires the number of function evaluations of a classical
algorithm working on a solution space reduced in size because one bit of $%
\mathbf{k}$, either $k_{0}$ or $k_{1}$, is known in advance. To identify the
missing bit, the classical algorithm has to perform just one function
evaluation -- for example, if the advanced information is $k_{0}=0$\ , it
should compute $\delta \left( \mathbf{k},x\right) $ for either $x=00$ (if $%
\delta \left( \mathbf{k},00\right) =1$ then $k_{1}=0$ if $\delta \left( 
\mathbf{k},00\right) =0$ then $k_{1}=1$ ) or $x=01$ (if $\delta \left( 
\mathbf{k},01\right) =1$ then $k_{1}=1$ if $\delta \left( \mathbf{k}%
,01\right) =0$ then $k_{1}=0$ ). This verifies the 50\% rule for $n=2$.

More in general, a classical algorithm that knows in advance 50\% of the $n$%
\ bits that specify the data base location, in order to identify the $n/2$\
missing bits should perform $O\left( 2^{n/2}\right) $ function evaluations,
against the $O\left( 2^{n}\right) $ of a classical algorithm without
advanced information. This verifies the 50\% rule for $n>2$.

This rule says that unstructured data base search is solvable with a
quadratic speed up independently of our knowledge of the quantum algorithm.
The speed up comes by comparing two classical algorithms, with and without
the advanced information.

\subsection{Building the quantum algorithm out of the advanced information
classical algorithm}

The function evaluation stage of the quantum algorithm -- namely the
transformation of $\Psi _{0}$ (equation \ref{preparation}) into $\Psi _{1}$
(equation \ref{secondstage}) -- is the sum of the histories of the advanced
information classical algorithm (see section 2.3).

The choice of the initial phase of each history is justified as in Deutsch's
algorithm.

We examine the outcome of the function evaluation stage, namely $\Psi _{1}$
(equation \ref{secondstage}). Registers $K$ and $X$ are maximally entangled,
orthogonal states of $K$, each corresponding to a value of $\mathbf{k}$, are
correlated with orthogonal states of $X$.\ This means that the value of $%
\mathbf{k}$\ has propagated to register $X$. To read this value, we should
rotate the $X$ basis in such a way that $-\left\vert 00\right\rangle
_{X}+\left\vert 01\right\rangle _{X}+\left\vert 10\right\rangle
_{X}+\left\vert 11\right\rangle _{X}$ (correlated with $\left\vert
00\right\rangle _{K}$) goes into $\left\vert 00\right\rangle _{X}$, etc.
This is a constructive definition of the transformation $U$.

Generalizing to $n>2$ is straightforward. Given the advanced knowledge of $%
n/2$ bits, in order to compute the missing $n/2$ bits we should perform
function evaluation and rotate the basis of $X$ an\ $\func{O}\left( 2^{\frac{%
n}{2}}\right) $ times. The first time we obtain a superposition of an
unentangled state of the form (\ref{preparation}) (the initial state
transformed into itself with a slightly smaller amplitude)\ and a maximally
entangled state of the form (\ref{final}). At each successive iteration, the
amplitude of the latter state is amplified at the expense of the amplitude
of the former, until it becomes about $1$ in $\func{O}\left( 2^{\frac{n}{2}%
}\right) $ iterations.

\section{Conclusions}

We have verified that the 50\% rule, the fact that a quantum algorithm
requires the number of function evaluations of a classical algorithm that
knows in advance 50\% of the information that specifies the solution of the
problem, holds for the main quantum algorithms.

This rule, besides shading light on the nature of quantum computation,
brings the search of the problems solvable with a quantum speed up to an
entirely classical framework. The quantum speed up comes out by comparing
two classical algorithms, with and without the advanced information. In
hindsight, one can see that the existing speed ups are skillfully designed
around the 50\% rule. This rule can be used for a systematic exploration of
the possibility of achieving speed ups, perhaps to explain why the speed ups
discovered until now are so few. Once identified a problem solvable with a
speed up, the same rule can be used a second time for the search of the
quantum algorithm that solves the problem. In fact the advanced information
classical algorithm defines the quantum algorithm.

The 50\% rule establishes a correspondence between quantum computation and
classical computation with advanced information. It is natural to ask
ourselves whether, in some more general sense, quantum mechanics is
classical mechanics with advanced information. This question would deserve
further investigation.

{\Large Acknowledgments}

Thanks are due, for encouragement and useful comments, to Vint Cerf, Artur
Ekert, David Finkelstein, Lov Grover, G\"{u}nter Mahler, and Hartmut Neven.

{\Large Bibliography}

Benioff, P. (1982). Quantum mechanical Hamiltonian models of Turing machines.%
\textit{\ J. Stat. Phys}., 29, p. 515.

Bennett, C.H. (1973). Logical reversibility of computation.\textit{\ IBM J.
Res. Dev.} 6, p. 525.

Bennett, C.H. (1982). The Thermodynamics of Computation -- a Review. \textit{%
Int. J. Theor. Phys.} 21, p. 905.

Castagnoli, G., and Finkelstein, D. (2001). Theory of the quantum speed up. 
\textit{Proc. Roy. Soc. Lond}. A 457, p. 1799. quant-ph/0010081.

Castagnoli, G. (2008). The mechanism of quantum computation. \textit{Int. J.
Theor. Phys., }vol. 47, number 8/August, 2008, p. 2181.

Castagnoli, G. (2009).\ The quantum speed up as advanced cognition of the
solution. \textit{Int. J. Theor. Phys}., vol. 48, issue 3, p. 857.

Castagnoli, G. (2009).\ The quantum speed up as advanced knowledge of the
solution. http://arxiv.org/pdf/0809.4545

Cleve, R., Ekert, A., Macchiavello, C., and Mosca, M. (1998). Quantum
Algorithms Revisited. Proc. Roy. Soc. Lond. A, vol. 454, number 1969, p. 339.

Deutsch, D. (1985). Quantum theory, the Church-Turing principle and the
universal quantum computer. \textit{Proc. Roy. Soc}. (Lond.) A, 400, p. 97.

Deutsch, D. and Jozsa, R. (1992). Rapid solution of problems by quantum
computation. \textit{Proc. Roy. Soc}. (Lond.) A, 439, p. 553.

Finkelstein, D. R. (1969). Space-Time Structure in High Energy Interactions.
Coral Gables Conference on Fundamental Interactions at High Energy. Center
of Theoretical Studies January 22-24, 1969. University of Miami. Timm
Gudehus, Geoffrey Kaiser, and Arnold Perlmutter Eds. Gordon and Breach,
Science Publishers, New York London Paris. pp. 324-343. //\ Finkelstein, D.
R. (1969). Space-time code. \textit{Phys. Rev.} 184, p. 1261.

Fredkin, E. and Toffoli, T. (1982). Conservative logic.\textit{\ Int. J.
Theor. Phys.} 21, p. 219.

Grover, L. K. (1996). A fast quantum mechanical algorithm for data base
search. Proc. 28th Ann. ACM Symp. Theory of Computing.

Kaye, P., Laflamme, R., and Mosca, M. (2007). An Introduction to Quantum
Computing. Oxford University Press Inc., New York.

Mosca, M. and Ekert, A. (1999). The Hidden Subgroup Problem and Eigenvalue
Estimation on a Quantum Computer. \textit{Lecture Notes in Computer Science}%
, Volume 1509.

Simon, D. (1994). On the Power of Quantum Computation. \textit{Proc. 35th
Ann. Symp. on Foundations of Comp. Sci. p.} 116.

\end{document}